# Intensity-resolved ionization yields of aniline with femtosecond laser pulses

J. Strohaber*, T. Mohamed, N. Hart, F. Zhu, R. Nava, F. Pham, A. A. Kolomenskii, H. Schroeder, G. G. Paulus, and H. A. Schuessler.

*Texas A&M University, Department of Physics, College Station, TX 77843-4242, USA*

*Corresponding author: jstroha1@physics.tamu.edu

We present experimental results for the ionization of aniline and benzene molecules subjected to intense ultrashort laser pulses. Measured parent molecular ions yields were obtained using a recently developed technique capable of three-dimensional imaging of ion distributions within the focus of a laser beam. By selecting ions originating from the central region of the focus, where the spatial intensity distribution is nearly uniform, volumetric-free intensity-dependent ionization yields were obtained. The measured data revealed a previously unseen resonant-like multiphoton ionization process. Comparison of benzene, aniline and Xe ion yields demonstrate that the observed intensity dependent structures are not due to geometric artifacts in the focus. Finally we attribute the ionization of aniline to a stepwise process going through the $\pi\sigma^*$ state which sits 3 photons above the ground state and 2 photons below the continuum.

PACS numbers: 32.80.Rm, 32.80.Wr, 4185.Ew, 4230.Wb

The interaction of intense and ultrashort laser pulses with molecules has generated considerable interest in the scientific community over the past few decades. Because of the complexities inherent to molecular systems, ultrashort intense-field ionization experiments tend to exhibit a more vibrant and richer behavior than those found for atoms. It is widely known that the many degrees of freedom and time scales associated with molecules present both an analytical and numerical challenge for theoreticians when calculating ionization rates of molecules [1—3]. As pointed out by S. L. Chin and co-workers, experimental and theoretical results from the interaction of ultrashort pulses with simpler systems (viz., benzene) may pave the way to a better understanding of more complex systems [3,4]; in other words, ionization dynamics from these systems may carry over to more complex systems which exhibit similar physical properties.

Previously, this line of thought has resulted in the intense investigation of the noble gases and diatomic molecules. A well-known example is found in the ionization of Ar and $N_2$ and in the ionization of Xe and $O_2$, where each pair shares similar ionization energies [5,6]. Investigations have shown that the ionization dynamics of $N_2$ were atomiclike, while those of $O_2$ were nonatomiclike. Oxygen exhibited suppressed ionization and this ionization suppression has been attributed to molecular symmetry. Generalizations have also been made in the investigation of electron trapping in noble gas atoms and in the CO molecule [7]. Analogously, in the current work, we find that aniline (amino-benzene) molecules having a phenyl ring exhibit similar ionization dynamics with the prototypical aromatic molecule benzene. By generalizing upon the ionization mechanism recently put forward for benzene [8,9], we conclude that the frontier orbitals, in addition to a low lying 3s-Rydberg state, are responsible for the observed nonlinearities in the ion yield curves.

In experiments, accurate measurements of ionization probabilities as a function of a well-defined intensity are hindered by the broad range of intensities within a laser beam. Spatial averaging is known to camouflage true ionization probabilities by smoothing out subtle features that may have been present in presaturation probabilities and to conceal all physical information in postsaturation yields [8,10]. Recently, it has been shown using a time-of-flight (TOF) spectrometer that direct measurements of ionization probabilities—previously thought close to impossible [12]—are obtainable. For the research presented here, the technique outlined in [11] has allowed for the acquisition of benzene and aniline parent molecular ion yields closely resembling "true ionization probabilities". There exist other methods to retrieve ionization probabilities (references found in Ref [11]); however, with the exception of the recently reported ion microscope [13], none have demonstrated the *direct* measurement of ionization probabilities.

Benzene has been studied extensively over the past few decades, and the photodynamics of the $S_0 \rightarrow S_1$ resonances near 266 nm in these molecules has been well established [14—17]. This wavelength conveniently coincides with that of frequency-quadrupled radiation ($1064 \text{ nm}/4 = 266 \text{ nm}$) from a Nd:YAG laser, and has facilitated earlier research on the photophysics of these molecules. In this previous research, ultraviolet nanosecond pulses were used to ionize benzene molecules in a (1+1) REMPI (pump-probe) scheme going through the $S_1$ state with data recorded as a function of laser wavelength [17]. In contrast, the current work investigates the ionization of benzene and aniline under the interaction of single infrared femtosecond pulse and as a function of laser intensity. We used ~45-fs, 800-nm pulses, and because $800 \text{ nm}/3 = 266 \text{ nm}$, the phenyl $S_1$ state in benzene can be accessed with 3 photons. The inherently larger bandwidth (in our case ~23 nm) of femtosecond pulses makes our experiment less sensitive to exact excitation energies of the $S_1$ intermediate level than in earlier nanosecond work. Another distinctive feature of the present research is found in the high optical pumping rates of ultrashort pulses. Compared to nanosecond pulses, femtosecond pulses promote efficient parent ion formation of the molecules by exceeding the dissociation rates and allowing the resulting parent ions to be identified with a TOF spectrometer [15]. To the best of our knowledge, intensity-resolved ion yields as a function of laser intensity and for ultrashort, 800-nm pulses have only been reported for benzene [8, 9], but not for aniline. We will demonstrate why the setups used in previous experiments could not have shown the now revealed resonant-like multiphoton ionization (MPI) aspect of these molecules under the interaction with ultrashort pulses.

For a given peak laser intensity $I_0$, experimentally measured ion yields $S(I_0)$ are the result of integrating weighted ionization probability $P(I)$ over all intensities within the focus,

$$S(I_0) \propto \int_0^{I_0} P(I) \left| \frac{\partial V(I, I_0)}{\partial I} \right| dI \quad . \tag{1}$$

Because of the volumetric weighting factor $\partial V / \partial I$ in Eq. (1), the measured ion signal $S(I_0)$ differs from the ionization probability $P(I)$. This is known in the literature as spatial averaging and is responsible for the 3/2-power law following saturation. A number of experimental and theoretical techniques have been devised to circumvent spatial averaging. These methods involve deconvolving experimental data by using mathematical algorithms [8], ion microscopy [11, 13 and refs therein] or a combination of both these methods [8,10].

To detect ions in the focus, we use a reflectron-type TOF ion mass spectrometer (modeled after the device described in Ref [11]) Fig. 1. By carefully adjusting its voltage settings, to control the TOF dispersion curve Figs. 1(a) and 1(b), this setup is capable of three-dimensional spatial imaging. Two of the dimensions of our detection volume were defined by the entrance slit sizes $\sim 10~\mu m$ by $\sim 400~\mu m$. By detuning our ion mirror, a one-to-one mapping between space and TOF Fig. 1(b) defined the remaining dimension of our detection volume $\sim 3~\mu m$. Because the maximum intensity of the laser was not needed, the detection volume was positioned at $z \approx 1.7~mm$ downstream to the focus. It can be shown that the spatial intensity variation ($\text{RSTD} = \sqrt{\sigma/\langle I^2 \rangle}$) within this volume is roughly 7%, which is on par with power fluctuations from our laser.

Our laser setup consisted of an 800-nm Ti:sapphire laser system delivering ~45-fs pulses at a repetition rate of 1kHz and a maximum average power output of ~1.0 W. Linearly polarized (along the TOF axis) laser radiation was focused into an evacuated ionization chamber using a 22.7 cm focal length lens. Ions were detected with a multichannel-plate detector and subsequently counted using a Fast Comtec multiscalar having a 250 ps time resolution. Typical background pressure in the ionization chamber was $9 \times 10^{-9}$ mbar, which was achieved after baking the apparatus. Contaminants in our sample were removed by vacuum distillation. We observed that the absence of contaminants went hand-in-hand with a stable operating pressure. For all measurements the total pressure was adjusted (after distillation) to $7.5 \times 10^{-7}$ mbar, since previous experiments with Xe ions indicated that in this pressure range space charge effects are negligible [11].

Figure 2 shows measured data for the aniline (blue squares) and benzene (red circles) parent molecular ions as a function of laser intensity. Since $n^{th}$-order MPI ion yields plotted in a log-log representation have slopes of $n$, it is commonly practiced to place straight line segments on the data to determine nonlinearities Fig. 2(a). For benzene, the ion yields at lower intensities ( $I_0 < 3\times 10^{13}$ W/cm$^2$ ) exhibit a slope of ~6. This low-intensity behavior is as expected from lowest-order perturbation theory since at least six photons ( $6\cdot 1.55$ eV $= 9.3$ eV ) are needed to ionize benzene ( IE $= 9.24$ eV ) from the ground state. A remarkable observation of the present work is the appearance of a "kink" or an abrupt change in slope as a function of laser intensity. In the intensity range of $3\times 10^{-13}$ W/cm$^2$ to $8\times 10^{-13}$ W/cm$^2$ our data deviates from that of Ref. [3] showing an intensity dependence $\propto I_0^3$. For intensities greater than $8\times 10^{-13}$ W/cm$^2$ a decrease in the ion yields is observed. This decrease is a telltale sign of the elimination of volumetric-weighting in intensity-resolved ionization experiments due to the increasing role of higher-order processes such as fragmentation and/or MPI of the target [8, 11, 13]. In our experiments, analysis of the benzene fragments showed that fragmentation is responsible for the postsaturation decrease but not the decrease following the kink. Only minimal amounts of multiply charged benzene parent ions were observed in the TOF spectrum.

In the case of aniline (blue squares), the ionization yield curves show a similar intensity-dependence to that of benzene, *i.e.,* a kink. The straight line segments shown in Fig. 2(a) indicate that the observed slopes for the aniline data are ~5 for the lower-intensity yields and ~2 for intensities preceding saturation but greater than the kink intensity. For ion yields taken over a range of intensity that include saturation effects, yields following an $I^n$ dependence will *not* hold for the entire yield curve, and differentiating between ion yields due to a "rollover to saturation" and those due to a genuine change in slope may be difficult to discern from plots of the data. For this reason, the data has been fitted to a function capable of supporting two slopes before saturation, which then goes to unity for all intensities thereafter. The fits are shown in the inset of Fig. 2(a). For benzene the slopes were found to be 6.4 for the lower intensities and 2.8 in the region directly preceding saturation and those of aniline were found to be 5.2 and 2.3 respectively. To further illustrate the presence of the kinks, the derivatives of the fit functions are shown. The presence of the kinks is indicated by plateaus.

The solid black curves in Fig 2(b) demonstrates why conventional TOF measurements could not have revealed the observed kink structures and are the result of spatial averaging our data over the Gaussian focal volume. [8]. In other words, our data was used as the theoretical ionization probability $P(I)$ in Eq. (1). This spatially averaged data is in agreement with experimental and theoretical ionization yields in Refs [3,4,5]. In our measurements, the dimension ($x$) of our detection volume created by mapping space into time-of-flight was decreased until the slope after the kink converged to the reported minimum value demonstrating that larger values of the slope may be found if the detection volume is not suitably chosen (residual spatial averaging).

Ionization measurements obtained using this type of detection method have only recently been demonstrated in imaging experiments. To determine if restriction of the focal volume (volumetric-clipping) is responsible for the observed kinks in the measured data, we performed computer simulations to investigate the effect of a finite detection volume within a Gaussian focus. We were unable to artificially produce kink structures using our experimental parameters. To investigate volumetric-clipping experimentally, we simultaneously measured ion yields of benzene and xenon (red circles and black triangles in Fig 2). For benzene, we observed the previously found ~6 to ~3 nonlinearities, while for Xe the intensity dependence was found to 4.8. Despite Xe requiring 8 photons to ionize, this ~5 scaling is consistent with previous experimental and theoretical (PPT) results [18]. We conclude from the absence of a kink in the Xe data that the observed intensity-dependencies in the ion yields of benzene and aniline are the indication of a genuine physical phenomenon.

The observation of integer power behavior suggests that the ionization mechanism is MPI rather than tunneling ionization (TI). The Keldysh parameter $\gamma = \sqrt{IE/2U_p}$ [19] for both benzene and aniline (where IE is the ionization energy and $U_p = 9.33 \times 10^{-14} I[\text{W/cm}^2]\lambda^2[\mu\text{m}]$ is the ponderomotive potential) is equal to unity for an intensity of $I \sim 10^{14}$. Tunneling is therefore expected to dominate for intensities much greater than this value. For benzene and aniline molecules, the ion yields were found to saturate at an intensity less than $10^{14}$ W/cm$^2$, therefore, MPI is expected to be the dominant ionization mechanism for intensities less than the observed saturation intensities.

For benzene, a proposed ionization mechanism has recently been reported in the literature [8,9]. This ionization mechanism was attributed to the three-photon excitation of the $S_1$ intermediate state, which readily saturated at higher intensities and is followed by ionization into the continuum. The rate-limiting step is from the intermediate state to the continuum and is therefore responsible for the slope following the kink. Additionally, dynamic effects such as the AC-Stark shift can be neglected because the $S_1$ state population occurs earlier within the pulse evolution and both the $S_1$ and continuum shift by similar amounts Fig 3. The ionization energy of aniline is 7.72 eV requiring a total of 5 photons ($5 \cdot 1.55$ eV $= 7.75$ eV) to ionize neutral aniline molecules from the $S_0$ state. Aniline was specifically chosen in our experiments because the odd number of photons needed to ionize aniline breaks the (3+3) symmetry in the number of photons needed for the ionization of benzene through its intermediate state. Because the slope after the kink was measured to be ~2, this suggests from the benzene results that an intermediate state exists, which lies two photons below the continuum. Theoretical calculations have shown that the primary $S_0 \rightarrow S_1$ transition in aniline is a HOMO-LUMO transition with a contribution from the HOMO-1 to LUMO +1, and the $S_0 \rightarrow S_2$ transition consist of a HOMO to LUMO+1 and a HOMO-1 to LUMO excitation; the relevant molecular orbits for these $\pi\pi^*$ transitions can be found in Ref [20]. The vertical excitation energy of the $S_0 \rightarrow S_1$ in aniline is 4.22 eV [20], which lies far outside the spectrum of the laser radiation. A plausible ionization channel is that the $S_1$ state ponderomotively shifts into a 3-photon resonance with the ground state. However, the continuum is also expected to shift by roughly the same amount and 3 photons would be needed to further ionize this excited state Fig 3, but we measure a nonlinearity of 2. The vertical transition energy of the $S_2$ state has been found experimental to be ~5.19 eV [20]. This ionization path would require 2 photons to further ionize into the continuum, but would require more than 3 photons to populate.

Based on the previous discussion, neither the $S_1$ nor $S_2$ $\pi\pi^*$ states alone seem to account for the observed kink in aniline. A state that lies at near resonance with the excitation energy of three photons 4.65 eV has recently been observed in TOF experiments using a (2+2) REMPI scheme [21, 22] and in total kinetic energy release TKER spectra using 269 nm radiation [23]. This so-called New "S2" state has a vertical excitation energy of 4.6 eV and is close to the excitation energy of three 800nm photons. This $\pi\sigma^*$ state is a low lying 3s-Rydberg state which sits between the $\pi\pi^*$ $S_1$ and $S_2$ states. These states are known to be populated either by direct photoexcitation or by radiationless transition from optically "bright" $\pi\pi^*$ states such as the $S_1$ and $S_2$ [24]. REMPI experiments have demonstrated a short vibronic structure in the $\pi\sigma^*$ state and theoretical calculations have verified a potential barrier for short N-H and C-N bond lengths. From these considerations, our experimental results indicate that the stepwise ionization of aniline proceeds through the recently discovered New "S2".

In summary, by subjecting aniline molecules to 800-nm, ~45-fs pulses, a REMPI process was revealed using an imaging spectrometer. By comparing ion yields of benzene and aniline to those of Xe we demonstrate that the observed kink structures—which may turn out to be ubiquitous feature in many molecules—are due to a genuine physical phenomenon. By using a recently proposed ionization mechanism for benzene [8,9], aniline was found to ionize primarily through a recently reported $\pi\sigma^*$ state.

This work was partially supported by the Robert A. Welch Foundation (No. A1546), MURI DOD (No.W911NF-07-1-0475), and the National Science Foundation (Nos. 0722800 and 0555568).

# REFERENCES

**References with titles and full author list**

1) S. Chelkowski, C. Foisy, A. D. Bandrauk, "Electron-nuclear dynamics of multiphoton dissociative ionization in intense laser fields," Phys. Rev. A **57,** 1176 (1998).

2) M. J. Dewitt and R. Levis, "Concerning the ionization of large polyatomic molecules with intense ultrafast lasers," J. Chem. Phys. **110,** 11368 (1999).

3) A. Talebpour, A. D. Bandrauk, K. Vijayalakshmi, and S. L. Chin, "Dissociative ionization of benzene in intense ultra-fast laser pulses," J. Phys. B: At. Mol. Opt. Phys. **33,** 4615 (2000).

4) R. Itakura, J. Watanabe, A. Hishikawa, and K. Yamanouchi, "Ionization and fragmentation dynamics of benzene in intense laser fields by tandem mass spectroscopy," J. Chem. Phys. **114,** 5598 (2001).

5) K. Nagaya, H.-F. Lu, H. Mineo, K. Mishima, M. Hayashi, and S. H. Lin, "Theoretical studies on tunneling ionizations from the doubly degenerate highest occupied molecular orbitals of benzene in intense laser fields," J. Chem. Phys. **126,** 024304 (2007).

6) C. Guo, M. Li, J. P. Nibarger, and G. N. Gibson, "Single and double ionization of diatomic molecules in strong laser fields," Phys. Rev. A **58,** R4271 (1998).

7) A. Talebpour, Y. Liang, and S. L. Chin, "Population trapping in the CO molecule," J. Phys. B: At. Mol. Opt. Phys. **29,** 3435 (1996).

8) J. Strohaber, A. A. Kolomenskii, and H. A. Schuessler, "Reconstruction of ionization probabilities from spatially averaged data in N dimensions," Phys. Rev. A **82,** 013403 (2010).

9) T. Scarborough, J. Strohaber, D. B. Foot, C. J. McAcy and C. J.G.J. Uiterwaal, "Ultrafast REMPI in benzene and the monohalobenzenes without the focal volume effect," Phys. Chem. Chem. Phys. **13,** 13783-13790 (2011).

10) P. Hansch, L. D. Van Woerkom, "High-precision intensity-selective observation of multiphoton ionization: a new method of photoelectron spectroscopy," Opt. Lett. **21,** 1286 (1996).

11) J. Strohaber, C. J. G. J. Uiterwaal, "In Situ Measurement of Three-Dimensional Ion Densities in Focused Femtosecond Pulses," Phys. Rev. Lett. **100,** 023002 (2008).

12) M. P. de Boer, J. H. Hoogenraad, R. B. Vrijen, R. C. Constantinescu, L. D. Noordam, and H. G. Muller , "Adiabatic stabilization against photoionization: An experimental study," Phys. Rev. A **50,** 4085 (1994).

13) M. Schultze, B. Bergues, H. Schroder, F. Krausz, and K. L. Kompa, "Spatially resolved measurement of ionization yields in the focus of an intense laser pulse," New J. Phys. **13,** 033001 (2011).

14) U. Boesl, H. J. Neusser, and E. W. Schlag. "Isotope selective soft multiphoton ionization and fragmentation of polyatomic molecules," J. Am. Chem. Soc. **103,** 5058 (1981).

15) R. J. Longfellow, D. B. Moss, and C. S. Parmenter, "Rovibrational level mixing below and within the channel three region of S1 benzene," J. Phys. Chem. **92,** 5438 (1988).

16) E. Loginov, A. Braun, and M. Drabbels, "A new sensitive detection scheme for helium nanodroplet isolation spectroscopy: application to benzene," Phys. Chem. Chem. Phys. **10,** 6107 (2008).

17) L. Zandee and R. B. Bernstein, "Resonance-enhanced multiphoton ionization and fragmentation of molecular beams: NO, I2, benzene, and butadiene," J. Chem. Phys. **71,** 1359 (1979).

18) A. Talebpour, C. Y. Chien, S. L. Chin, "Coulomb effect in multiphoton ionization of rare-gas atoms," J. Phys. B: At. Mol. Opt. **31,** 1215 (1998).

19) M. J. Nandor, M. A. Walker, L. D. Van Woerkom, "Angular distribution of high-intensity ATI and the onset of the plateau," J. Phys. B: At. Mol. Opt. Phys. **31,** 4617 (1998).

20) E. Drougas, J. G. Philis, and A. M. Komas, "Ab initio study of the structure of aniline in the $S_1$ and $S_2$ $\pi\pi^*$ states," J. Mol. Struc-THEOCHEM **758,** 17-20 (2006).

21) Y. Honda, M. Hada, M. Ehara, and H. Nakatsuji, "Excited and ionized states of aniline: symmetry adapted cluster configuration interaction theoretical study," J. Chem. Phys. **117,** 2045 (2002).

22) T. Ebata, C. Minejima, and N. Mikami, "A new electronic state of aniline observed in the transient IR absorption spectrum from S1 in a supersonic jet," J. Phys. Chem. A **106,** 11070 (2002).

23) G. A. King. T. A. A. Oliver, and M. N. R. Ashfold, "Dynamical insight into state mediated photodissociation of aniline," J. Chem. Phys. **132,** 214307 (2010).

24) M. N. R. Ashfold, *et. al.,* J. Strohaber, D. B. Foot, C. J. McAcy and C. J.G.J. Uiterwaal, " $\pi\sigma^*$ excited states in molecular photochemistry," Phys. Chem. Chem. Phys. **12,** 1218-1238 (2009).

**References without titles and shortened author list**

1) S. Chelkowski, C. Foisy, A. D. Bandrauk, Phys. Rev. A **57,** 1176 (1998).

2) M. J. Dewitt and R. Levis, J. Chem. Phys. **110,** 11368 (1999).

3) A. Talebpour, A. D. Bandrauk, K. Vijayalakshmi, and S. L. Chin, J. Phys. B: At. Mol. Opt. Phys. **33,** 4615 (2000).

4) R. Itakura, J. Watanabe, A. Hishikawa, and K. Yamanouchi, J. Chem. Phys. **114,** 5598 (2001).

5) K. Nagaya, H.-F. Lu, H. Mineo, K. Mishima, M. Hayashi, and S. H. Lin, J. Chem. Phys. **126,** 024304 (2007).

6) C. Guo, M. Li, J. P. Nibarger, and G. N. Gibson, Phys. Rev. A **58,** R4271 (1998).

7) A. Talebpour, Y. Liang, and S. L. Chin, J. Phys. B: At. Mol. Opt. Phys. **29,** 3435 (1996).

8) T. Scarborough, J. Strohaber, D. B. Foot, C. J. McAcy and C. J.G.J. Uiterwaal, Phys. Chem. Chem. Phys. **13,** 13783-13790 (2011).

9) J. Strohaber, A. A. Kolomenskii, and H. A. Schuessler, Phys. Rev. A **82,** 013403 (2010).

10) P. Hansch, L. D. Van Woerkom, Opt. Lett. **21,** 1286 (1996).

11) J. Strohaber, C. J. G. J. Uiterwaal, Phys. Rev. Lett. **100,** 023002 (2008).

12) M. P. de Boer, J. H. Hoogenraad, R. B. Vrijen, R. C. Constantinescu, L. D. Noordam, and H. G. Muller, Phys. Rev. A **50,** 4085 (1994).

13) M. Schultze, B. Bergues, H. Schroder, F. Krausz, and K. L. Kompa, New J. Phys. **13,** 033001 (2011).

14) U. Boesl, H. J. Neusser, and E. W. Schlag, J. Am. Chem. Soc. **103,** 5058 (1981).

15) R. J. Longfellow, D. B. Moss, and C. S. Parmenter, J. Phys. Chem. **92,** 5438 (1988).

16) E. Loginov, A. Braun, and M. Drabbels, Phys. Chem. Chem. Phys. **10,** 6107 (2008).

17) L. Zandee and R. B. Bernstein, J. Chem. Phys. **71,** 1359 (1979).

18) A. Talebpour, C. Y. Chien, S. L. Chin, J. Phys. B: At. Mol. Opt. **31,** 1215 (1998).

19) M. J. Nandor, M. A. Walker, L. D. Van Woerkom, J. Phys. B: At. Mol. Opt. Phys. **31,** 4617 (1998).

20) E. Drougas, J. G. Philis, and A. M. Komas, J. Mol. Struc-THEOCHEM **758,** 17-20 (2006).

21) Y. Honda, M. Hada, M. Ehara, and H. Nakatsuji, J. Chem. Phys. **117,** 2045 (2002).

22) T. Ebata, C. Minejima, and N. Mikami, J. Phys. Chem. A **106,** 11070 (2002).

23) G. A. King. T. A. A. Oliver, and M. N. R. Ashfold, J. Chem. Phys. **132,** 214307 (2010).

24) M. N. R. Ashfold, *et. al.,* J. Strohaber, D. B. Foot, C. J. McAcy and C. J.G.J. Uiterwaal, Phys. Chem. Chem. Phys. **12,** 1218-1238 (2009).

**Figure Captions**

Fig. 1. (color online) Schematic view of the 3D imaging spectrometer. TOF apparatus: G1-G3 grid-electrodes of the ion mirror, (SP) slit plate, (RP) repelling plate, (MCP) multichannel plate. (Blue peanut shape) isointensity shell near the focus. (Red parallelepiped) detection volume within the focus. (a) Time-of-flight as a function of position between the slit and repeller plates. (b) One-to-one mapping of distance to arrival time for a slightly detuned reflectron (imaging mode).

Fig. 2. Intensity dependent ion yields: ion yields of aniline (blue squares), benzene (black circles) and xenon (red triangles); (a) ion yields of aniline exhibit a slope of 5 to 2, ion yields of benzene have a slope of 6 to 3 and those of xenon do not show a change in slope. (b) integrated ion yields over a focal volume demonstrate why conventional TOF setups could not have revealed the resonant like behavior in the ion yields curves for aniline and benzene.

Fig. 3. Time dependent energy level diagrams: (a) During the leading edge of the pulse, three 800-nm photons excite the mainly unperturbed benzene molecule from the ground state $S_0$ to the singlet $S_1$ state. The intermediate state may then be ionized later during the same pulse; (b). Ionization mechanism proposed for benzene also hold for aniline including the New "S2" state between the $S_1$ and $S_2$ states.

**Figures**

Figure 1

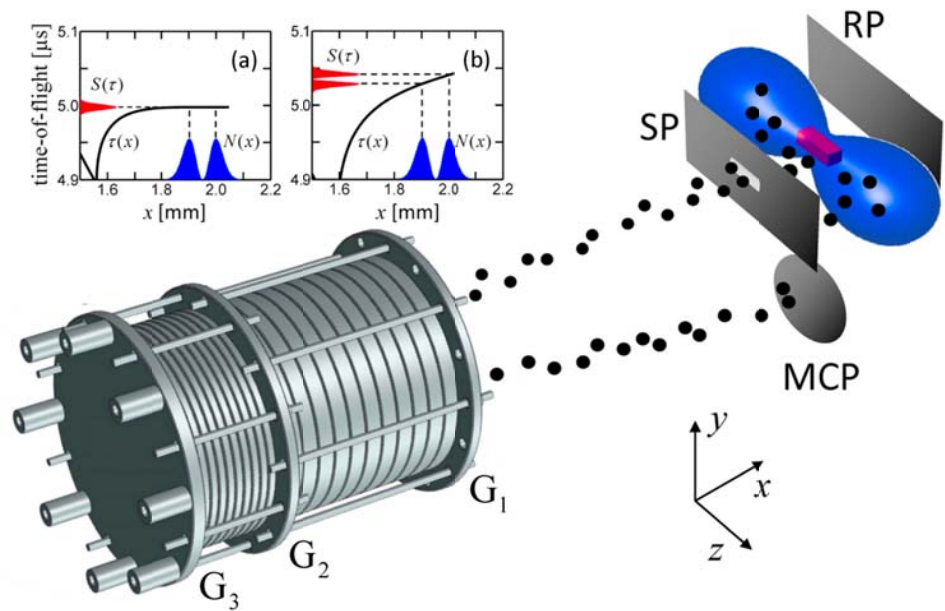

Figure 2

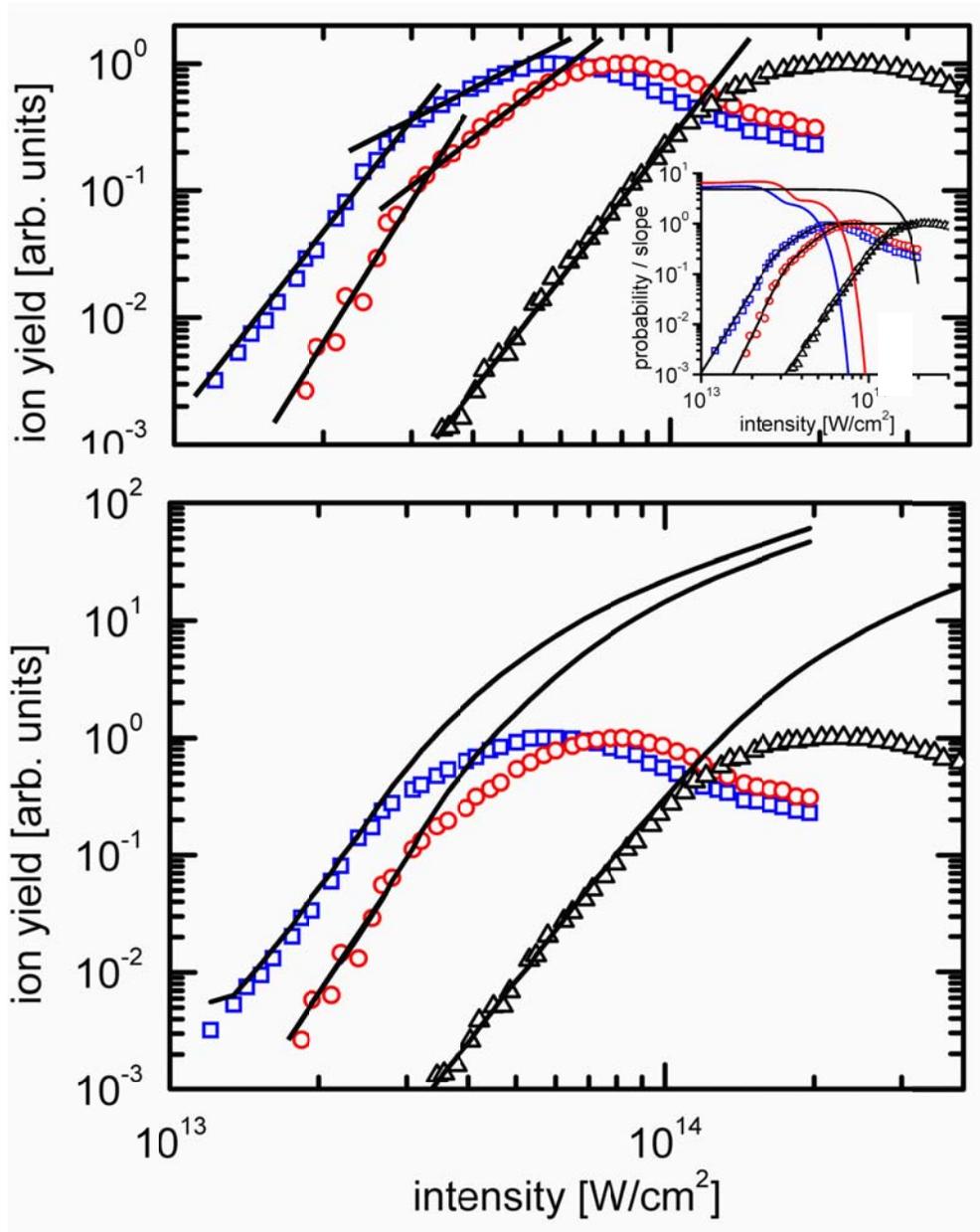

Figure 3

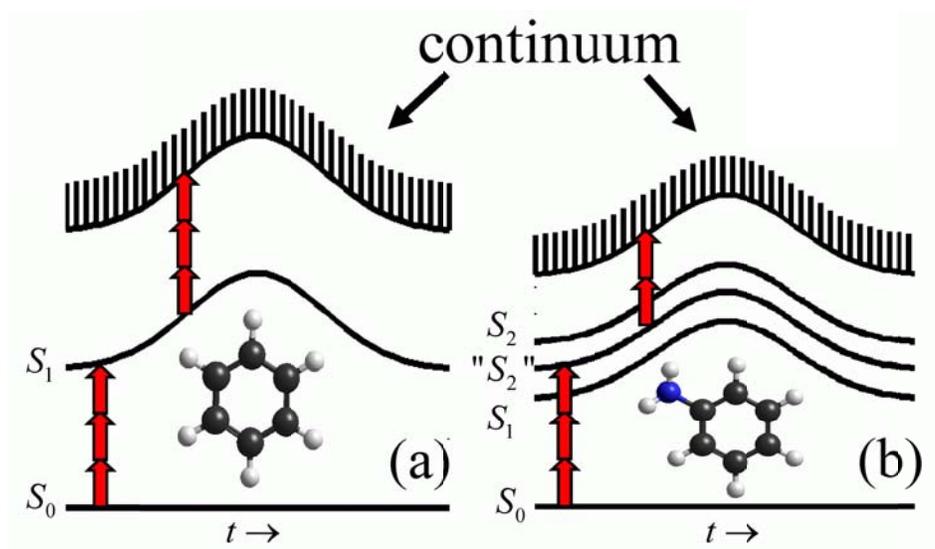